\documentclass[aps,prl,twocolumn]{revtex4}
\usepackage{graphicx}
\usepackage{amssymb}
\usepackage{latexsym,amsmath,amsthm}
\usepackage[dvips]{color}
\newcommand{\gaa}{g_{aa}}
\newcommand{\gac}{g_{ac}}
\newcommand{\ta}{t_a}
\newcommand{\tc}{t_c}
\newcommand{\bosnum}{N_a}
\newcommand{\D}{\Delta(x)}

\newcommand{\hnja}{{\hat n}_j^a}
\newcommand{\hnjc}{{\hat n}_j^c}
\newcommand{\nia}{n_i^a}
\newcommand{\nja}{n_j^a}
\newcommand{\njab}{{\bar n}_j^a}
\newcommand{\epsq}{\varepsilon_q}
\newcommand{\rtm}{u_{n}(x)}
\newcommand{\ltm}{v_{n}(x)}
\newcommand{\psio}{\psi_{1} (x)}
\newcommand{\psit}{\psi_{2} (x)}
\newcommand{\vf}{v_F}
\newcommand{\pot}{V(x)}

\newcommand{\alpa}{1/2}
\newcommand{\inalpa}{2}

\newcommand{\hc}{{\hat c}}
\newcommand{\ha}{{\hat a}}
\newcommand{\hd}{{\hat d}}

\newcommand{\bm}{{\bar m}}

\begin{document}
\title {Holstein model and Peierls instability in 1D
boson-fermion lattice gases}

\author{E. Pazy and A. Vardi}

\affiliation{Department of Chemistry, Ben-Gurion University of the
Negev, P.O.B. 653, Beer-Sheva 84105, Israel}

\date{\today}

\begin{abstract}
We study an ultracold bose-fermi mixture in a one dimensional 
optical lattice. When boson atoms are heavier then fermion atoms 
the system is described by an adiabatic Holstein model, exhibiting 
a Peierls instability for commensurate fermion filling factors. 
A Bosonic density wave with a wavenumber of twice the Fermi 
wavenumber will appear in the quasi one-dimensional system.  
\end{abstract}

\maketitle

The recent realization of Bose-Fermi mixtures of ultracold 
atoms \cite{Truscott01,Hadzibabic01,Schreck01,Roati02} is 
very promising for studying various strong correlation phenomena,
with Bose fields replacing the lattice phonons of classic condensed-matter 
models. Virtual exchange of the bose-field excitations were
shown to induce fermion-fermion attractive interactions 
\cite{Heiselberg00,Bijlsma00,Modugno02,Simoni03}, leading 
to Cooper-like pairing \cite{Efremov02,Viverit02,Matera03}
and enhancing the transition to fermion superfluidity.
Furthermore, completely novel phenomena were predicted,
such as the formation of boson-fermion composite fermionic 
particles \cite{Kagan,Lewenstein04,Sanpera04} and their subsequent 
pairing into quartets \cite{Kagan}.

One particularly interesting direction is the
possibility of mutual fermion-boson trapping in 
optical lattices \cite{Lewenstein04,Sanpera04,Roth04,
Buchler03,Mathey04}, opening the way to the realization 
of various discrete lattice models. For electrons confined 
to a quasi-one-dimensional electric conductor
the coupling to bosonic phonons leads to the Peierls instability
towards a charge density wave (CDW) with wave number equal to 
twice the Fermi wavelength $2k_F$. The corresponding 
instability for a Bose-Fermi mixture in an harmonic trap
was recently predicted \cite{Miyakawa04}. In this article we 
study the Peierls instability in a mixture of light fermionic and 
heavy bosonic atoms in a quasi-one-dimensional optical 
lattice, shown to be described by an adiabatic Holstein model. The 
resulting CDW depicted schematically in Fig. 1, consists of 
both a fermionic density wave and a spatial modulation in the 
bosonic density, with twice the Fermi wave length. Fermionic 
atoms and bosonic modulations will either position in alternate 
sites (Fig. 1a) or in the same sites (Fig. 1b) depending on the 
sign of the fermion-boson interactions.   
    
\begin{figure}
\centering
\includegraphics[scale=0.45]{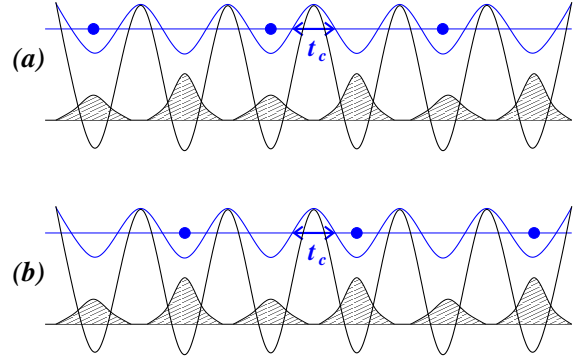}
\caption{Peierls instability in a lattice fermi-bose mixture: (a) repulsive
fermion-boson interactions (b) attractive fermion-boson interactions.
The shaded part depicts the bosonic mean field density whereas filled circles
denote fermionic atoms.}
\label{fig:potential}
\end{figure}

We consider a mixture of $N_c$ spin polarized 
fermionic atoms and $\bosnum$ bosonic atoms
confined to a quasi one-dimensional (Q1D) optical lattice 
with $M$ sites (see Fig. \ref{fig:potential}). For a strong 
optical field, one can expand boson- and fermion field operators 
in terms of the one-mode-per-site Wannier basis set \cite{Jaksch},
thus obtaining the lowest Bloch-band Hubbard model,  
\begin{eqnarray}
H &=& - \sum_{\langle jk \rangle} \tc \hc_j^\dag 
\hc_k +  \gac\sum_{j} \hnjc \hnja
+ \frac{\gaa}{2}\sum_j \hnja(\hnja-1) \nonumber \\
&-& \sum_{\langle jk \rangle} \ta \ha_j^\dag\ha_k+\frac{\omega_0^2\ell^2}{2}
\sum_{j} j^2
\left (m_c\hnjc+ m_a\hnja\right)  \ ,
\label{eq:Hamiltonian}
\end{eqnarray}
where the operators $\hc_j$ and $\ha_j$ annihilate a spin polarized fermion 
and a boson respectively, in the $j$-th site. The density operators 
$\hnjc=\hc_{j}^{\dagger} \hc_{j}$ and $\hnja=\ha_{j}^{\dagger}\ha_{j}$ are 
the fermionic and bosonic densities respectively. The fermion and
boson atomic masses and hopping amplitudes are $m_c,t_c$ and $m_a,t_a$ 
respectively. The collisional terms $\gaa$ and $\gac$ correspond to 
on-site boson-boson interaction which will be assumed positive (repulsive) throughout the paper, and onsite fermion-boson interaction, respectively.  The last term on the r.h.s. of Eq.~\ref{eq:Hamiltonian}
is the harmonic trap potential  where $\omega_0$, is the relevant oscillator 
frequency and $\ell$ is the lattice spacing.

In the weak-interaction limit the bosonic field in Eq. (\ref{eq:Hamiltonian}) 
may be treated in a mean-field approximation, replacing the bosonic 
density $\hnja$ by its $c$-number expectation value 
$\nja=\langle \ha_j^{\dagger}\ha_j\rangle$. 
We will further assume that 
the bosonic atoms are heavier than the fermionic atoms 
as for a $^{6}$Li-$^{87}$Rb mixture.  
Since the tunneling term depends exponentially 
on the atomic mass and since the 
dynamic polarizability of the larger boson atoms is greater than
the polarizability of the fermions leading to effectively deeper
traps for the bosons (see Fig. 1), we can neglect the bosonic hopping 
term and retain only fermion tunneling \cite{overlap}. In this limit
the system can be described by an adiabatic Holstein model, where 
its ground state is found adiabatically by solving
the 'fast' fermionic problem
\begin{equation}
H_{eff}^c 
(\{\nia\})= - \sum_{\langle jk \rangle} \tc \hc_{j}^{\dagger} 
\hc_{k} + \sum_{j} \left(\frac{m_c \omega_0^2\ell^2}{2}j^2+\gac\nja\right)\hnjc  \ ,
\label{eq:effHamc}
\end{equation}
treating the bosonic densities $\nja$ as fixed parameters and then adding  
the resulting fermion energy (which depends parametrically on $\{\nja\}$) 
as an effective potential to the 'slow' bosonic
Hamiltonian,
\begin{equation} 
H_{eff}^a=\frac{\gaa}{2}\sum_j {(\nja)}^{2}+ \frac{m_a\omega_0^2\ell^2}{2}\sum_{j} j^2\nja~.
\label{eq:effHama}
\end{equation}

For $\omega_0=0$, the adiabatic Holstein model is known to exhibit a Peierls 
instability \cite{Peierls}, with respect to bosonic
collective excitations with wave number equal to $N_c/M$,
corresponding to twice the Fermi wave length $k_F=N_c/2M$ \cite{Peierls2}. 
It is favorable for the system to reduce the 1D translation symmetry 
by enlarging the effective unit cell, thereby opening a gap in 
the fermionic spectrum. When the wavelength of the excitation is 
$2k_F$ this gap coincides with the discontinuity of the Fermi 
distribution, so that all fermions are on the side
of the spectrum which lowers in energy. For example, for $N_c/M=1/2$ 
the unit cell doubles, opening a gap in the fermionic spectrum at 
the zone boundary of the folded Brillouin zone. It should be noted that 
in difference to the standard Su-Schrieffer-Heeger (SSH)
model \cite{SSH} used for quasi one-dimensional systems 
exhibiting a  Peierls instability,
the coupling to the bosonic degrees of freedom 
in our system is on-site whereas in the SSH model it effects the 
hopping probability. 

\begin{figure}
\centering
\includegraphics[scale=0.50]{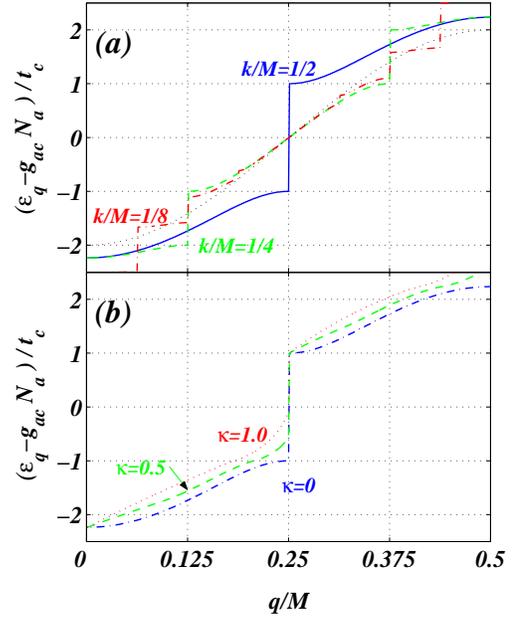}
\caption{Fermionic energy spectrum for various values of the bosonic
modulation wavenumber with $\omega_0=0$ (a) and as a function of 
$\omega_0$ for a fixed modulation with $k/M=1/2$ (b).}
\label{fig:spectrum}
\end{figure}
 
In order to demonstrate the Peierls instability we will study 
how the energy of the system is affected by spatial bosonic modulations 
of the form 
\begin{equation}
\nja=\njab+\delta n_j^a\cos{(\frac{2\pi k j}{M})} \ .
\label{eq:orderpara}
\end{equation} 
The density $\njab=[\mu-(m_a \omega_0^2 j^2/2)]/\gaa$ with $\mu$ denoting the 
chemical potential of the bosons, is the Thomas-Fermi density
profile which minimizes the fixed $N_a$ bosonic energy 
$H^a_{eff}+\mu(\sum_j \hnja-N_a)$, in the 
absence of fermion-boson interactions. The density modulation depth 
$\delta n_j^a \ll\njab$ is 
generally a function of $j$, varying slowly compared to the modulation
wavelength. 

\begin{figure}
\centering
\includegraphics[scale=0.45]{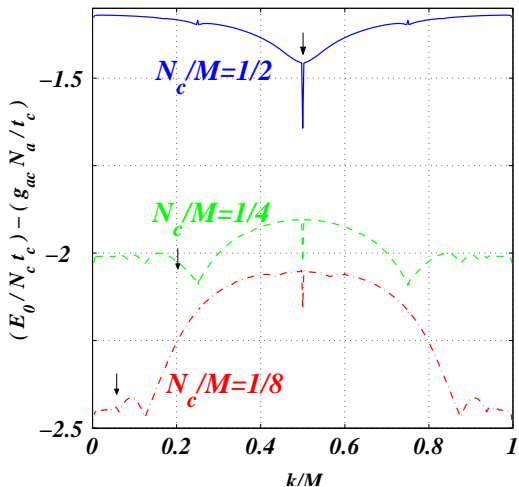}
\caption{Ground-state fermionic energy as a function of the 
modulation wavenumber for various fermion filling factors: 
$N_c/M=1/2$ (solid), $N_c/M=1/4$ (dashed), $N_c/M=1/8$ (dash-dotted).
The external trap frequency is set to $\kappa=0.1t_c$ 
and the bosonic amplitude modulation is set equal to $t_c$. Arrows
indicate the bosonic modulation wavenumber minimizing $E_c$.}
\label{fig:fermionicEn}
\end{figure}

Under the ansatz (\ref{eq:orderpara}) the fermion Hamiltonian 
(\ref{eq:effHamc}) takes the form
\begin{eqnarray}
H_{eff}^c&=& - \sum_{\langle jk \rangle} \tc \hc_{j}^{\dagger} 
\hc_{k} + \frac{\gac}{\gaa}\mu \sum_j \hnjc\nonumber\\
~&~&+\sum_{j} \left[\kappa\left(\frac{j}{M}\right)^2
+\Delta_j\cos{\left(2\pi k\frac{j}{M}\right)}\right]\hnjc  \ ,
\label{eq:effHamct}
\end{eqnarray}
where $\kappa=\bm(\omega_0\ell M)^2/2$ with 
$\bm=m_c-(\gac/\gaa)m_a$, and $\Delta_j=\gac\delta n_j^a$.
Evidently, the mutual trapping of both fermions and bosons can
only take place when $\gac/\gaa<m_c/m_a$ or fermion 
atoms will scatter out of the trap by the bose mean-field.
In what follows we shall assume that this condition is satisfied.
In Fig. 2 we plot the fermionic energy spectrum,
$\varepsilon_q$, as a function  of the Fermionic wave number,
$q$, obtained from
direct diagonalization of the fermionic Hamiltonian (\ref{eq:effHamct}) 
for constant $\Delta_j=\Delta$. In Fig. 2a 
the harmonic trap frequency is set equal to zero, whereas the effect of the 
trap is demonstrated in Fig. 2b by fixing the modulation frequency to $k=M/2$
and plotting the spectrum for various values of $\omega_0$. The bosonic modulation distorts
the periodicity of the lattice, thereby opening a gap at $q=k/2$. For 
$k/M=2k_F$ the gap coincides with the Fermi momentum, 
so that all the states with $|q|/M<k/2M=k_F$ whose energy is lowered are full 
and all the states with $|q|/M>k/2M=k_F$ which increase in energy are empty. 
Consequently, the fermionic energy is minimized for $k=2k_F$, as depicted
in Fig. \ref{fig:fermionicEn} where we plot the fermionic ground-state 
energy $E_c$ obtained by integration over the fermion spectrum up to the 
Fermi energy, as a function of the wavenumber of the spatial modulation 
in the boson field. Sharp minima are attained as expected, for 
$k/M=N_c/M=2k_F$. Further local minima of the energy, corresponding to
smaller gaps opening at the Fermi momentum, also appear for integer 
multiples of $k/M$ .   

\begin{figure}
\centering
\includegraphics[scale=0.45]{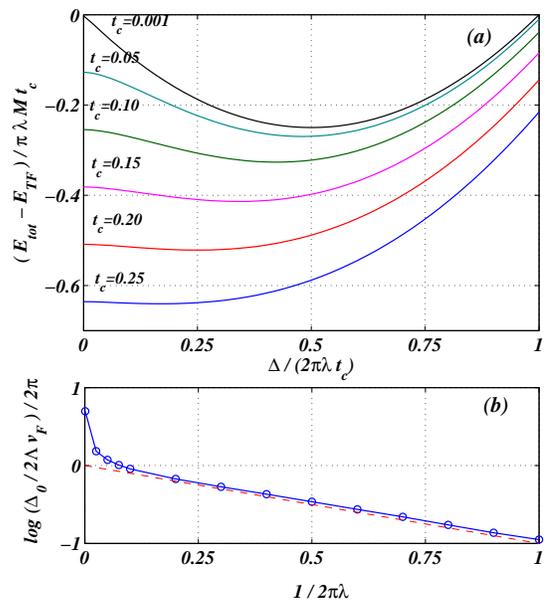}
\caption{Total energy of the system as a function of the gap $\Delta$
for various values of $t_c$ (a) and optimal gap values as a function
of $\tc$ (b). The dashed line depicts the predicted strong coupling 
behavior of Eq. (\ref{eq:gap}).}
\label{fig:Etot}
\end{figure}

The total energy of a half-filled system with $k=M/2$ 
is plotted in Fig. \ref{fig:Etot}a, as a function of the 
modulation depth $\Delta$.
As can be seen from the proceeding analysis, the reduction 
in $E_c$ due to the bosonic modulation is approximately linear 
in  $\Delta$. By substitution of 
(\ref{eq:orderpara}) into the bosonic Hamiltonian (\ref{eq:effHama}) 
we see that the boson contribution to the total energy is 
$E_a=E_{TF}+(\gaa/2\gac)\sum_j \Delta_j^2 \cos^2(2\pi k j /M)$
where $E_{TF}=(5/7)\mu N_a$ is the Thomas-Fermi energy. 
Thus, boson-boson repulsive interactions increase the energy 
quadratically with the modulation depth. Consequently there is always
a minimum in the total energy of the system $E_{tot}=E_c+E_a$ at some 
finite modulation amplitude indicating
that the formation of a CDW is indeed energetically favorable. The optimal 
modulation depth decreases as $\tc$ increases since linear fermionic 
dispersion is attained at decreasingly small values of the gap. The
resulting CDW can be detected by means of atom interferometry where the
doubling of the unit cell will be manifested in the inverse lattice
spacing or via the measurement of the collective excitation spectrum  
by means of Bragg spectroscopy \cite{Rey}.
 
Further insight into these numerical results is gained by 
employing the commonly used continuum model. For simplicity,
we will focus in what follows, on the half filling case 
$N_c/M=1/2$ where the 
bosonic order parameter minimizing $E_c$ is of the form
$n_j^a=\bar{n}_j^a+\delta n_j^a\cos(\pi j)$. 
We note that similar treatment can be applied for other
commensurate fermion filling factors. 
In the continuum limit, the fermionic Hamiltonian (\ref{eq:effHamct}) 
is rewritten as 
\begin{equation}
H_{c}=\int dx \Psi^{\dagger}(x)[-\frac{1}{2m}\frac{\partial^2}{\partial x^2}
\sigma_3 
+ \D \sigma_2 +  \pot \sigma_0]\Psi(x) \ ,
\label{eq:continuum}
\end{equation}
where    
\begin{eqnarray}
\Psi (x)\equiv \left (
\begin{array}{c}
\psio\nonumber \\
\psit
\end{array} \right )
\label{eq:vector}
\end{eqnarray}
is the spinor representation of the fermionic field in terms of 
right- and left-moving atoms, $\sigma_i$ are Pauli matrices
$\sigma_0$ is the identity matrix, and $m$ is the effective atomic mass.
The continuum limit for the trap potential is 
$\pot= m \omega_0^2 x^2/2$ and the gap parameter
is $\Delta_j\rightarrow\Delta(x)$. We note that in the Takayama-Lin-Liu-Maki
(TLM) model \cite{TLM} which is the continuum
limit of the SSH model, there is no confining potential 
and the dispersion is linearized. Moreover $\sigma_1$ appears for the 
coupling between left and right movers because the TLM coupling to 
phonons modifies the off-diagonal hopping rate, whereas in our case 
the coupling to the bose field modifies the diagonal self-energy 
terms. 

The fermion spectrum is obtained by Bogoliubov-de Gennes (BdG) 
diagonalization of the fermionic Hamiltonian. We follow
a similar method to the one used by Anderson to calculate the 
excitation spectrum of a superconductor with local disorder 
\cite{Anderson}. In this technique, which is essentially equivalent 
to a local density approximation (LDA), the fermionic spectrum 
is calculated by spatial averaging over spectra with different 
local order parameters. We expand the 
field operators $\psi_{1,2}(x)$ as
\begin{eqnarray}
\psio=\sum_{n}\rtm \hd_{n}&,&\psit=\sum_{n}\ltm \hd_{n},
\label{eq:expansion}
\end{eqnarray}
where $\hd_{n}$ are fermionic mode annihilation operators. The
functions $u_n(x),v_n(x)$ are assumed to take the form
\begin{eqnarray}
\rtm=\Phi_n(x)u_n&,&\ltm=\Phi_n(x)v_n,
\label{eq:spatialeigen}
\end{eqnarray}
where $\Phi_n(x)$ are harmonic oscillator eigenfunctions. Substituting
Eq. (\ref{eq:expansion}) and Eq.(\ref{eq:spatialeigen}) into Eq.
(\ref{eq:continuum}) and requiring that 
$H_c=\sum_{n}\varepsilon_{n} \hd_{n}^{\dagger} \hd_{n}$, we obtain 
two coupled BdG equations. In the framework of the LDA, we diagonalize 
the BdG equations for a local order parameter $\Delta$. 
The functions $\Phi_n(x)$ diagonalize the spatial part of the BdG 
equations which simplify to 
\begin{eqnarray}
(\varepsilon_n-E_n) u_n =-\imath \Delta v_n \nonumber \\
(\varepsilon_n+E_n) v_n= \imath \Delta u_n \ .
\label{eq:BdG}
\end{eqnarray}
Diagonalization of (\ref{eq:BdG}) results in the local
fermionic spectrum $\varepsilon_n=\sqrt{E_n^2+\Delta^2}$
in terms of the oscillator's energy $E_n=(n_F-n+1/2)\omega_0$
measured with respect to the Fermi energy, $E_{n_F}=(n_F+1/2)\omega_0$. 
This spectrum compares well with the numerical spectra of Fig 2b at the
continuum limit. In the limit $\omega_0 \rightarrow 0$, we have 
$n\omega_0 \rightarrow \vf q$,  where 
$\vf=2\tc\ell$ is the Fermi velocity (with $\hbar$ set equal to 1) and 
$q$ is the wave number
for the plain wave solution of the non-confined problem, so that 
the well known spectrum for a $\cos(\pi j)$ modulation 
$\epsq=\pm\sqrt{(q \vf)^2+\Delta^2}$
is reproduced. 

Having found the fermionic local spectrum, the total energy functional
of the system is given 
as the sum $E_{tot}=E_c+E_a$. The fermion energy $E_c$ is given 
within the LDA as
\begin{equation}
E_c=\sum_{n=1}^{N_c} \int dx | \Phi_n(x)|^2
\sqrt{ E_n^2+\Delta(x)^2} \simeq \sum_{n=1}^{N_c}\sqrt{ E_n^2+\Delta_0^2},
\label{eq:energy}
\end{equation}
where $\Delta_0$ is a constant order
parameter whose value is the spatial average of $\Delta(x)$. The boson contribution $E_a$ is given by 
\begin{equation}
E_a=\frac{1}{2\pi\lambda\vf}\int  \D^2 dx\ ,
\label{eq:boson}
\end{equation} 
where $\lambda={\gac}^2/(2\pi\gaa\tc)$, is the dimensionless fermion-boson
coupling constant. For a constant $\Delta(x)=\Delta_0$ we have 
$E_a=(\gaa/2\gac^2)M\Delta_0^2$. Minimizing $E_{tot}$ by setting its variation
with repect to $\D$ to zero, we obtain a 
self-consistent gap equation for $\D$,
\begin{equation}
\Delta(x) =  {\lambda \omega_0   \over 2}\sum_{n=1}^{N_c}| \Phi_n(x)|^2 
{\Delta_0 \over \sqrt{E_n^2+\Delta_0^2}} \,
\label{eq:selfconsistant}
\end{equation}
which is similar in form to the gap equation obtained by 
Anderson \cite{Anderson}. 

For sufficiently wide traps, $| \Phi_n(x)|^2$ can be replaced 
by its average value, thus restoring the familiar gap equation 
\begin{equation}
1=\vf \lambda \int_{0}^{\Lambda}{dq}
{\left  ( \sqrt{{(\vf q }^2+\Delta_{0}^{2}} \right )}^{-1},
\label{eq:selfcosistant}
\end{equation} 
where $\Lambda=\pi/2\ell$ is a momentum cut-off of the order of the 
fermionic band width. In the weak coupling regime 
$v_F \Lambda \ll \Delta_0$,  Eq. (\ref{eq:selfcosistant}) is solved
by $\Delta_0=\vf\lambda\Lambda=\gac^2/2\gaa$, whereas in the strong 
coupling regime $v_F \Lambda \gg \Delta_0$ we have the well known solution  
\begin{equation}
\Delta_0 = 2 v_F \Lambda \exp{(-1/\lambda)} \ .
\label{eq:gap}
\end{equation}
Our numerical results agree well with these continuum predictions as
demonstrated in Fig. 4. The weak coupling limit is confirmed by the 
low $\tc$ curves in Fig. 4a, which attain a minimum
at $\Delta_0=\pi\lambda\tc=\vf\lambda\Lambda$. The strong coupling
behavior is depicted in Fig. 4b where the minimum energy gap $\Delta_0$
is shown to precisely follow Eq. (\ref{eq:gap}) (dashed line) for 
sufficiently large $\tc$.  

The formation of a CDW in the Fermi-Bose mixture will be affected
by finite temperature effects. To observe Peierls instability
the temperature should be lower than the gap $T \ll \Delta_0 $. 
The fermionic excitation spectrum depends exponentially on the 
ratio $T /\Delta_0 $ so that the number of excited fermionic atoms
should be exponentially small. However the density of thermal bosonic
excitations has a power-law behavior $n_{ex}=(\bosnum/V){(T/T_c)}^{\alpa}$,
where $V$ is the volume and $T_c$ is the critical temperature.
This leads to a more restrictive constraint 
$T/T_c\ll {(\Delta_0/\gac\bar{n})}^{\inalpa}$, where $\bar{n}$ is
the bosonic density.

\bigskip

\begin{acknowledgments}
We gratefully acknowledge stimulating discussions with B. Horovitz. 
This work was supported in part by grants from the Minerva Foundation
through a grant for a Minerva Junior Research Group, the 
U.S.-Israel Binational Science Foundation (grant No.~2002214),
and the Israel Science Foundation for a Center of 
Excellence (grant No.~8006/03).
\end{acknowledgments}


\begin{thebibliography}{99}

\bibitem{Truscott01}
A. G. Truscott {\it et al.}, Science {\bf 291}, 2570 (2001).

\bibitem{Hadzibabic01}
Z. Hadzibabic {\it et al.}, Phys. Rev. Lett. {\bf 88}, 160401 (2001).

\bibitem{Schreck01}
F. Schreck {\it et al.}, Phys. Rev. Lett. {\bf 87}, 080403 (2001). 

\bibitem{Roati02}
G. Roati {\it et al.}, Phys. Rev. Lett. {\bf 89}, 150403 (2002). 

\bibitem{Heiselberg00}
H. Heiselberg {\it et al.}, Phys. Rev. Lett. {\bf 85}, 2418 (2000).

\bibitem{Bijlsma00}
N. J. Bijlsma {\it et al.}, Phys. Rev. A {\bf 61}, 053601 (2000).

\bibitem{Modugno02}
G. Modugno {\it et al.}, Science {\bf 297}, 2240 (2002).

\bibitem{Simoni03}
A. Simoni {\it et al.}, Phys. Rev. Lett. {\bf 90}, 163202 (2003).

\bibitem{Efremov02}
D.V. Efremov and L. Viverit, Phys. Rev. B {\bf 65}, 134519 (2002).

\bibitem{Viverit02}
L. Viverit, Phys. Rev. A {\bf 66}, 023605 (2002).

\bibitem{Matera03}
F. Matera, Phys. Rev A {\bf 68}, 043624 (2003).

\bibitem{Kagan}
M. Y. Kagan {\it et al.}, cond-mat/0209481.

\bibitem{Lewenstein04}
M. Lewenstein {\it et al.}, Phys. Rev. Lett. {\bf 92}, 050401 (2004).

\bibitem{Sanpera04}
A. Sanpera {\it et al.}, Phys. Rev. Lett. {\bf 93}, 040401 (2004). 

\bibitem{Roth04}
R. Roth and K. Burnett, Phys. Rev. A {\bf 69}, 021601 (2004).

\bibitem{Buchler03}
H. P. B\"uchler and G. Blatter, Phys. Rev. Lett. {\bf 91}, 130404 (2003).

\bibitem{Mathey04}
L. Mathey {\it et al.}, quant-ph/0401151.

\bibitem{Miyakawa04}
T. Miyakawa, H. Yabu and T. Suzuki, 
cond-mat/0401107.

\bibitem{Jaksch}
D. Jaksch {\it et al.}, Phys. Rev. Lett. {\bf 81}, 3108 (1998).

\bibitem{overlap}
To certain extent, the adiabatic limit in our case is 
even better justified than in the original Holstein model
which relies on the fixed electron/atom mass-ratio, since
$t_{a}/t_b$ is determined by the overlap of the Wannier 
functions, decaying exponentially with mass-ratio and relative
trap depths.


\bibitem{Peierls}
P. Peierls  {\em Quantum Theory of Solids}
,({Clarendon Press}, {Oxford}, 1955).

\bibitem{Peierls2}
G. Benfatto, G. Gentile and V. Mastropitro, 
J. Stat. Phys. {\bf 92}, 1071 (1998);
H. Fehske, A. P. Kampf ,M.  Sekania and G. Wellein 
Euro. Phys. J. B {\bf 31}, 11 (2003);
H. Fehske {\it et al.}, Phys. Rev. B {\bf 69},  165115 (2004).

\bibitem{SSH}
W. P. Su , J. R. schrieffer, and A. J. Heeger,  
Phys.  Rev. Lett. {\bf 42}, 1698 (1979).

\bibitem{Rey}
A. M. Rey {\it et al.}, cond-mat/0406552.

\bibitem{TLM}
H. Takayama, Y. R. Lin-Liu, K. Maki, Phys. Rev. B {\bf 22},  2099 (1980)
; {\it ibid}{\bf 28} 1138 (1983).

\bibitem{Anderson}
P. W. Anderson, J. Phys. Chem. Solids {\bf 11}, 26 (1959).

\end{thebibliography}
\end{document}